\documentclass[3p]{elsarticle}
\usepackage{amssymb,amsmath,mathtools}

\usepackage{graphicx}
\usepackage{cancel}
\usepackage[ruled,vlined]{algorithm2e}

\usepackage[colorinlistoftodos,textwidth=4cm,shadow]{todonotes}

\newcounter{Igor}

\date{}
\title {\textbf{A ``String Art'' Approach to the Design and Manufacturing of Optimal Composite Materials and Structures}}

\author{Igor Ostanin \footnote{Corresponding author, e-mail:i.ostanin@skoltech.ru}
}

\address{Center for Computational and Data-Intensive Science and Engineering, Skolkovo Institute of Science and Technology, Moscow, Russia.}

\begin{document}

\begin{abstract}

In this paper we report a new promising idea on the design and manufacturing of ply composite structures, tailored to exhibit maximum stiffness under given weight constraints and loading conditions. It is based on the idea behind an artistic technique known as ``string art'' - the representation of an image with a single thread, tensioned between pins on the flat frame. A discrete optimization algorithm has been employed recently to formalize the process of finding a configuration of thread windings that fuses into a given greyscale image. We demonstrate how this algorithm can be employed to approximate the two-dimensional distribution of isotropic material, computed by a conventional topology optimization algorithm. An optimal composite design is thus found as a result of the following two-stage process. At the first stage, topology optimization procedure produces regular grid of greyscale values of stiffness. At the second stage, this distribution is approximated with a polyline, representing the stiff reinforcement fiber in a soft matrix, according to ``string art'' optimization algorithm. The efficiency of the proposed approach is illustrated with few simple numerical examples. Our development opens a wide avenue for the industrial design of the new generation of fibrous composite structures. 

\end{abstract}

\maketitle

\section{Introduction}

The last decades have seen rapid development of the production technology of fiber-reinforced composites with superior properties. Modern carbon fiber-epoxy composite materials exceed the best aluminum alloys in terms of strength-to-weight ratio, which makes them indispensable for modern aerospace industry. Novel materials enable aircraft designs with outstanding characteristics, including high useful load and low fuel consumption. New generation of passenger jets employing novel composite materials allow for cheaper and greener commercial flights. Carbon nanotubes (CNTs) and other monocrystalline filaments promise even more impressive properties of prospective composites \cite{CNTcomp}. Another important area of application of such composite materials are civil engineering structures made of rod- and fiber-reinforced concrete \cite{reinf_concr}. 

Fiber-reinforced composites are typically based on fabrics or regular arrangements of strong and stiff fibers, embedded onto a relatively loose and compliant matrix. The combination of these two phases might be nearly as stiff and strong as the reinforcement fibers under tension, having much higher bending stiffness than the individual fibers.

The question of how exactly fibers should be laid into a matrix, in order to achieve the best possible mechanical performance of a composite part under a given set of loads, is highly non-trivial. Moreover, assuming that the optimal layout is found, it might be extremely difficult to produce it using the feasible manufacturing technology. 

In this work we suggest a possible remedy for both the issue of optimal design of fibrous composites and the issue of their manufacturability. Our idea relies on the following two-stage process. First, we find a greyscale material density distribution that minimizes the cost functional (\textit{e.g.} global strain energy), for a given boundary value problem using conventional Solid Isotropic Material with Penalization (SIMP) approach from the theory of topology optimization \cite{SIMP1,SIMP2,Sigmund99} (Fig.1(A,B)). Second, we try to find a good approximation of this material distribution in a basis of multiple straight fibers of constant thickness with arbitrary position and orientation. In order to solve this complex mathematical problem, we employ the approach known as ``string art'' - knitting of perceivable greyscale image with a multiple windings of a single thread, tensioned between multiple pins on the frame (Fig. 1(C)). This problem has been rigorously stated and solved in a recent work \cite{String1}. Placing the found configuration of fibers onto a soft matrix will result in a composite layer with stiffness optimized against the given loading configuration. 
In the following we describe our technique in details, solve few benchmark problems and discuss the important issues related to applicability of our approach and possible manufacturing technology of such composite plates. Concluding remarks section discusses possible generalizations of the presented approach.

\begin{figure}
	\begin{center}
	\includegraphics[width=16cm]{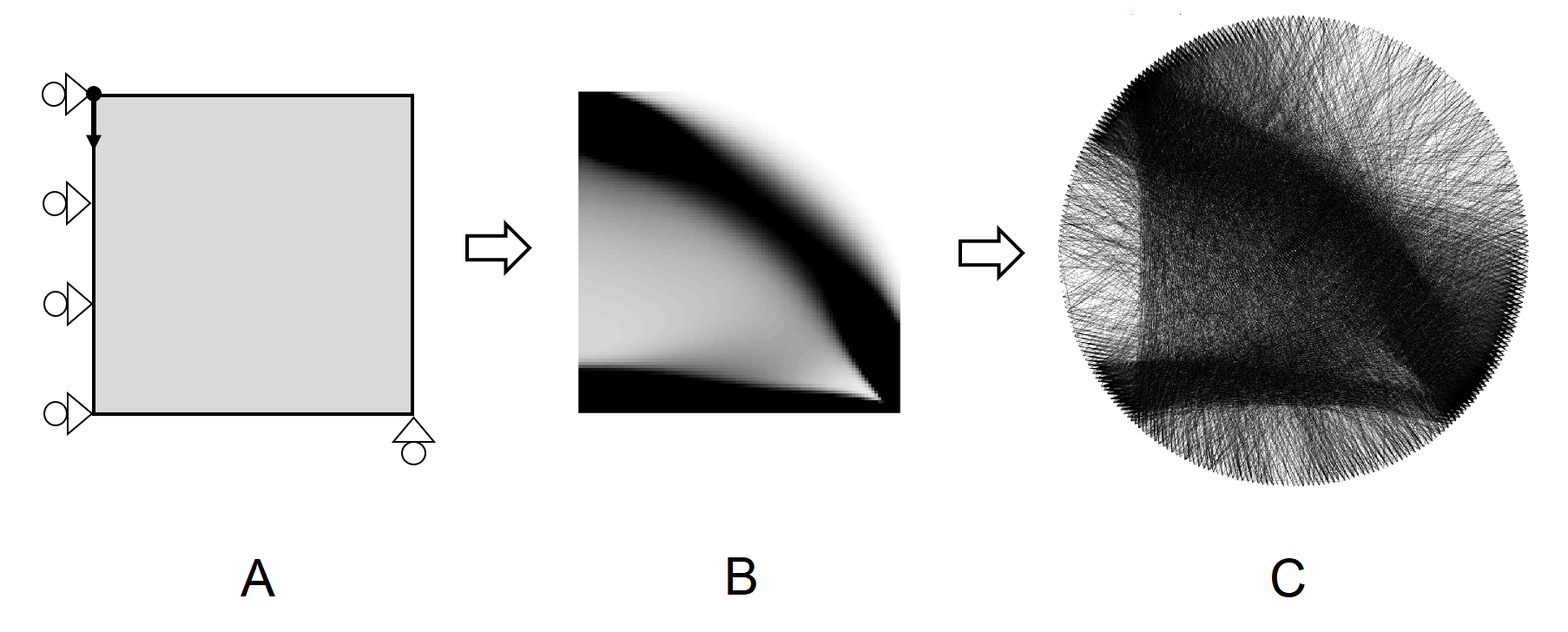}
	\protect\caption{Conceptual illustration of the proposed pipeline for manufacturing and design of an optimal composite structure (A) Initial boundary value problem, (B) material density distribution obtained by topology optimization algorithm. (C) ``String art'' approximation of the optimal density distribution}
	\end{center}
\end{figure}

\section{Method}

In order to present here a viable conceptual illustration of our technique, we use a rather simple setup. We start with the SIMP formulation of shape and topology optimization, that is used to find a globally optimal solution of a well-known compliance minimization problem in terms of greyscale distribution of linear isotropic material. This material distribution is then approximated in the basis of straight fibers of constant thickness embedded into a light an soft matrix. Then we solve a boundary value problem for the approximated problem geometry and evaluate the difference in the cost functional. Below we discuss the details of our numerical experiment.

\subsection{Topology optimization algorithm}

We employ standard MATLAB implementations of SIMP-based topology optimization codes based on optimality criteria, described in the works \cite{Sigmund99,Andreassen2011}. We start with a regular grid of 4-node square isoparameteric finite elements, with isotropic stiffness matrix and stiffness of the element controlled by design variable $\bf{x}$, assigned to every element, that will be further referred to as its density. We seek to find a vector of densities $\bf{x}$ that minimizes the given cost functional. In our example the cost functional is the global elastic strain energy - a measure of structure compliance. A SIMP formulation of the PDE-constrained compliance minimization problem can be written as follows:

\begin{eqnarray} \label{e1}
\min_{\mathbf{x}}: & & c(\mathbf{x})=\mathbf{U}^{T} \mathbf{KU}=\sum_{e=1}^{N} \mathbf{x}_{e}^{P} \mathbf{u}_{e}^{T} \mathbf{k}_{0} \mathbf{u}_{e}, \nonumber \\
\text{subject to:} & & \frac{V(\mathbf{x})}{V_0}=f, \nonumber \\
\text{:} & & \mathbf{KU=F}, \nonumber \\
\text{:} & & \mathbf{0 < x_{min} \leq x \leq 1}. \nonumber \\ 
\end{eqnarray}
  
Here $\bf{F}$ and $\bf{U}$ are global force and displacement vectors, respectively, $\bf{K}$ is global stiffness matrix, $\bf{u_e}$ and $\bf{k_e}$ are element displacements vector and stiffness matrix, respectively, $\bf{x_{min}}$ is a vector of minimum relative densities, $N$ is the number of finite elements used to discretize the design domain, $p$ is the penalization power, $V(\bf{x})$ and $V_0$ is the material volume and design domain volume, respectively and $f$ is the prescribed volume fraction.

The vector of minimum relative densities $\bf{x_{min}}$ is usually used to avoid the singularity in computations by using ersatz materials in ``empty'' areas.  

The optimality criteria algorithm as proposed by \cite{Bendsoe1995} is employed for gradient-based solution of this PDE-constrained optimization. We refer the reader to the detailed description of the algorithm in the educational article \cite{Sigmund99}; the code was modified for better performance in \cite{Andreassen2011}.

We have extracted FEM solvers used in topology optimization routine \cite{Andreassen2011} to compute the elastic fields and global strain energy of the arbitrary input distribution of material densities, given as the greyscale image.

\subsection{``String art'' approximation}

The algorithm (1) results in a greyscale distribution of the material (Fig.1(B)) that minimizes the cost functional. This complex material distribution needs to be approximated by certain realizable design, such that the elastic properties of this design are relatively close to the ones of an optimal material distribution. 

In this work we suggest a particular technology for production of such approximations, that is based on the idea known as ``string art'' in design and computer graphics communities (the credit belongs to P. Vrellis - see \cite{String1} and references therein). According to it, we approximate a given greyscale pixel distribution by a union of the background constant density and a polyline - multiple windings of a single continuous line of constant stiffness, tensioned around pins on a frame, as shown in Fig. 2(A, B). An evident advantage of this technology is its remarkable simplicity. In contrast with modern 3D printing techniques used for production of optimal parts, material distributions close to optimal can be created with the single degree of freedom machine, as illustrated in Fig. 2(C), which opens the wide avenue for a large-scale production of such composite structures.

It should be noted here that, strictly speaking, there are four distinct ways to connect two pins with a line, depending on the direction of winding (see the detailed discussion in \cite{String1}). In this work, we assume the thicknesses of pins and fibers negligible, leading to a unique way to connect every two pins with a string.

Few different algorithms have been developed for the best approximation of a given greyscale distribution with the set of wire windings \cite{String1}. They can be roughly divided onto ``greedy'' algorithms and ``global'' ones. ``Greedy'' algorithm computes each next pin joint based on the initial material distribution to be approximated and all the previous joints. ``Global'' algorithm i) computes the set of lines that best approximate the given pixel brightness distribution, ii) adds auxiliary joints to complete this set to a traversable Eulerian graph, and iii) forms a continuous path traversing the graph using Hierholzer algorithm (see the detailed discussion in \cite{String1}). In our work we utilize simple generator \cite{Generator}, that follows greedy strategy. Given brightness distribution $\mathbf{x}$, obtained by the topology optimization algorithm, we seek the sequence of pins $S$ that best approximates this distribution. In fact, we're solving the following discrete optimization problem:

\begin{equation} \label{e2}
\min_{\mathbf{y}}: d = (\mathbf{x} - F(\mathbf{y}))^2
\end{equation}
 
here $\mathbf{y}$ is the sequence of pins and $\mathbf{F(y)}$ is the corresponding set of pixel brightnesses (Fig.2(D) provides an illustration on how the mapping is performed). Within greedy approach, the problem \ref{e2} is solved according to the Algorithm \ref{a1}.  

\begin{algorithm}[H]\label{a1}
	\SetAlgoLined
	Initial thread position $y_0$

	Distribution $\mathbf{x}^r \coloneqq \mathbf{x}$
	
	\For{$i=1..N_{max}$}{
		compute $y_i$ s. t.: 
		
		$\min_{i}: (\mathbf{x}^r - f(y_i))^2$

		$\mathbf{x}^r \coloneqq \mathbf{x}^r - f(y_i)$
		
		if $(\mathbf{x}^r - f(y_i))^2 > (\mathbf{x}^r)^2$: stop
		
	}
	\caption{``Greedy'' algorithm to generate ``String Art'' pattern}
\end{algorithm}
\bigskip

In our case pixel brightnesses represent the stiffnesses of finite elements according to (1). The algorithm of drawing lines employed in our work \cite{Generator} utilizes anti-aliasing. It was earlier demonstrated \cite{Ostanin} that the anti-aliased line on a regular grid of finite elements represents well the properties of a straight elastic beam, even in the limit of fine lines.
 
It is important to note that in our formulation any crossed lines correspond to joint elastic beams. This corresponds to the situation of the composite material with cross-linked elastic fibers.   

It should be explicitly stated that in such formulation we do not solve the problem of finding optimal distribution of windings that mininizes elastic strain energy in the structure - as has been mentined above, the solution of such a formulation appears to be computationally intractable. Moreover, the algorithm \ref{a1} that we employ in this work does not guarantee even local minimization of \ref{e2}.   Instead, we use rather phenomenological observation that placing the reinforcement fibers at the locations that locally improve the functional \ref{e2}, we obtain a rather good approximation of the solution of \ref{e1} (see the numerical examples section).   

\begin{figure}
	\begin{center}
		\includegraphics[width=16cm]{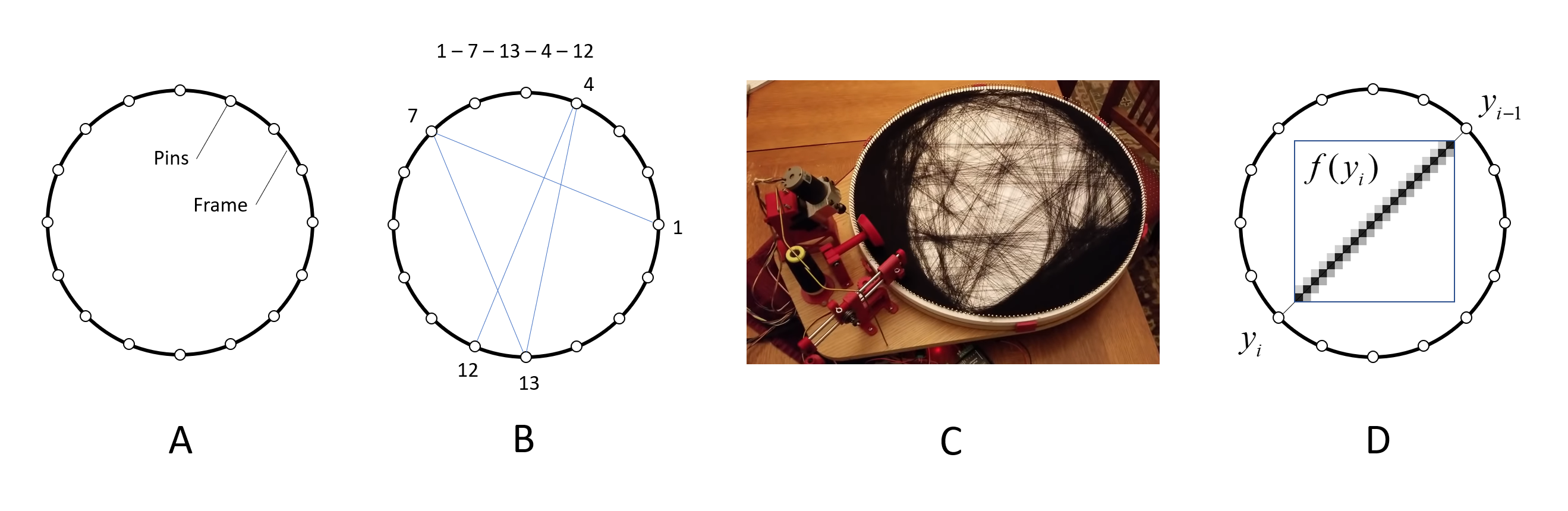}
		\protect\caption{``String art'' technology and its simulation. (A) ``String art'' frame (hoop) with evenly spaced pins. (B) String winding procedure illustration. (C) Single coordinate ``String art'' machine (Image credit - Barton Dring, printed with permission). (D) Conceptual illustration of mapping the line into a set of pixel brightnesses. }
	\end{center}
\end{figure}

\subsection{Simulation pipeline}

We employ the following simple scheme to illustrate the efficiency of the ``string art'' approach. We first generate the greyscale solution of the two-dimensional topology optimization problem of interest. Then, using the ``string art'' generator \cite{Generator}, we generate certain approximation of this greyscale distribution, using reasonable parameters of the generator (number of pins $N_p$, line brightness $\alpha$, number of lines $N_l$). We then balance the obtained approximation of the working area such that the total material mass constraint is satisfied. It is done in the following way. If the total mass of the fibers is higher than the mass constraint - the fiber brightness is decreased to meet the constraint. If the total mass is lower - the background brightness (mimicking the matrix material) is increased, to meet the mass constraint. Next, we measure relative $L^2$ difference between the original image and its ``string art '' approximation. Finally, the ``string art'' approximation for optimal material geometry, is treated as the distribution of elastic material (cross-linked fibers embedded into a soft matrix), and the total strain energy is computed. The obtained value of the strain energy is estimated against the minimum value reached by the optimal material distribution and the ``baseline'' value associated with the uniform material distribution.

\begin{figure}
	\begin{center}
		\includegraphics[width=16cm]{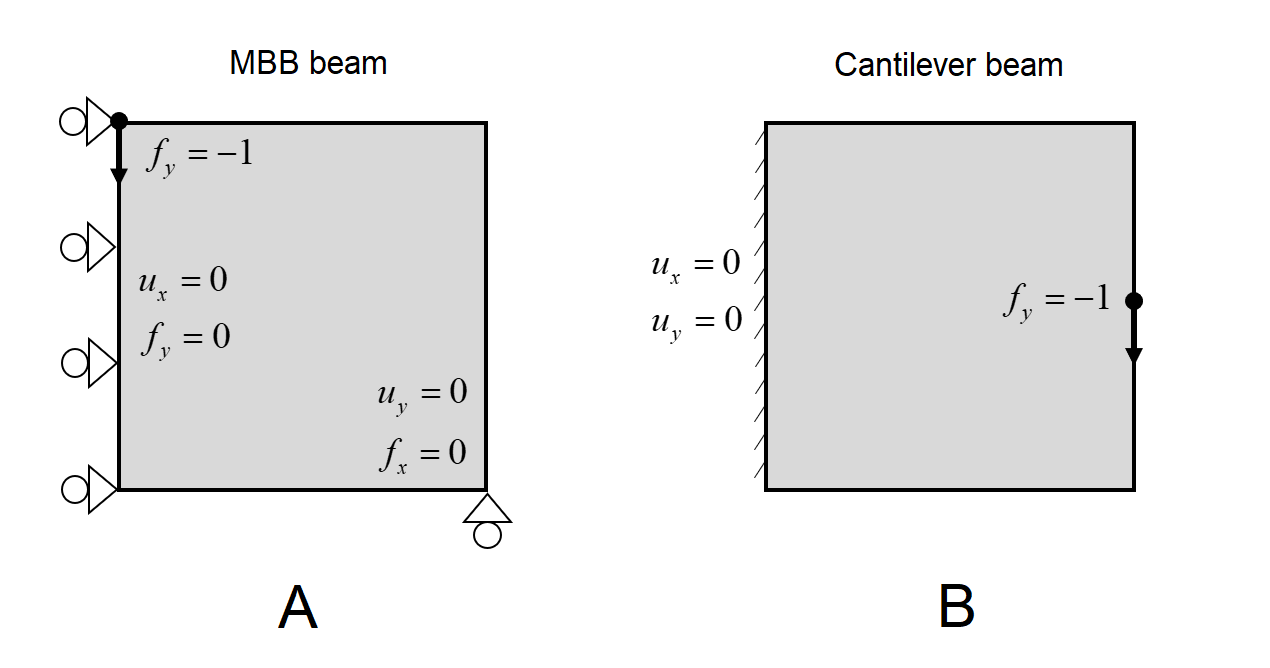}
		\protect\caption{Benchmark problems. (A) MBB beam, (B) Symmetric cantilever beam.}
	\end{center}
\end{figure}

\section{Numerical results} \label{examples}

In order to illustrate the possibility of using ``string art'' algorithms for efficient composite design, we provide here simple benchmark numerical examples, allowing to access how the original greedy algorithm of polyline approximation \ref{a1} approximates the classical SIMP solutions of topology optimization. We utilize online sting art generator \cite{Generator} and slight modifications of matlab FEM and topology optimization codes \cite{Sigmund99, Ostanin}.

\begin{figure}
	\begin{center}
		\includegraphics[width=16cm]{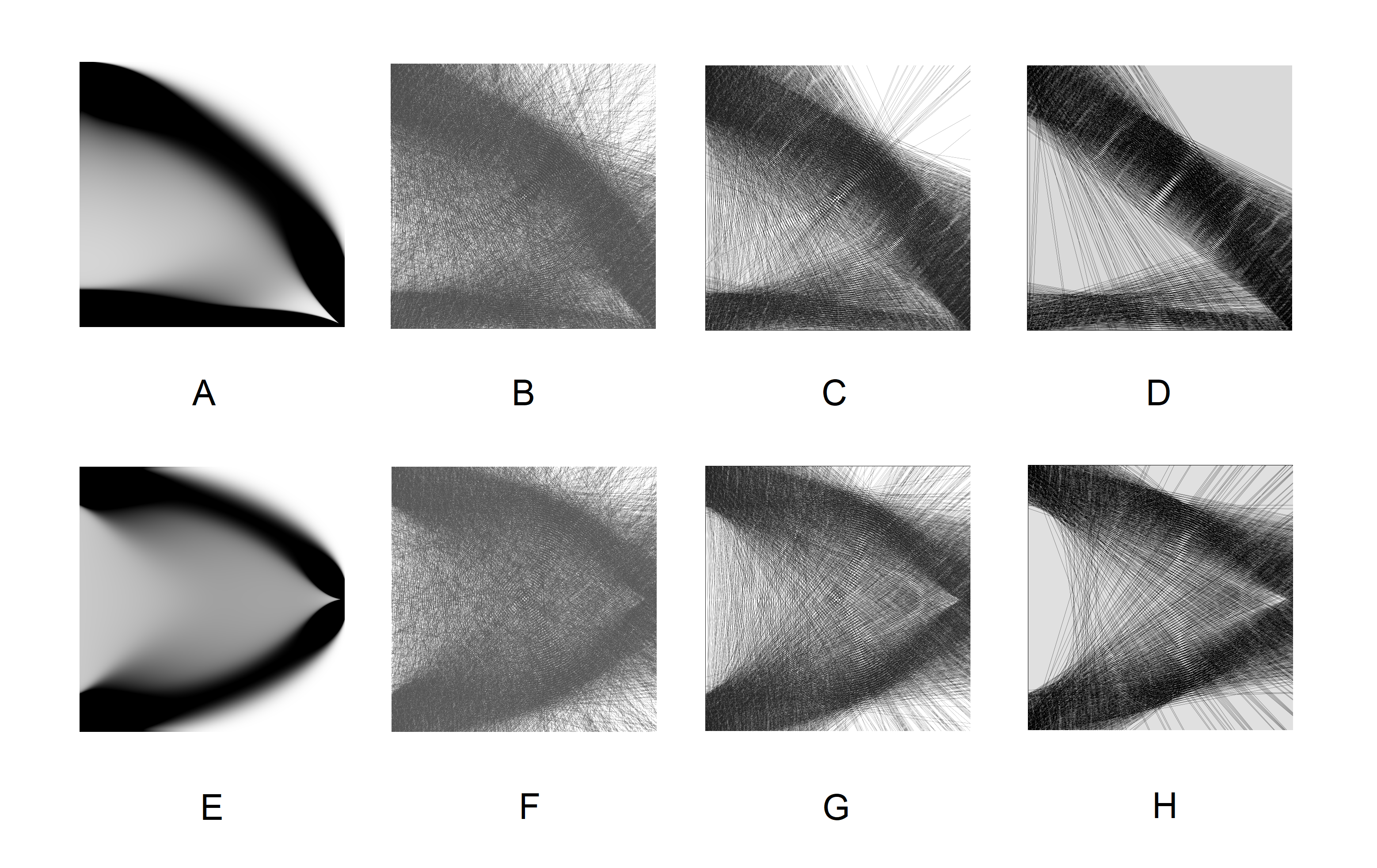}
		\protect\caption{Optimal solutions of benchmark problems and their ``string art'' approximations.}
	\end{center}	
\end{figure}

Consider two classical benchmark problems of a topology optimization - MBB beam (Fig. 3(A)) and symmetric cantilever beam (Fig. 3(B)), addressed by many authors and with a variety of methods (\textit{e.g.} \cite{Sigmund99, Andreassen2011}). This problem is first solved by the classical SIMP algorithm \ref{e1} without penalization ( $p=1$, globally optimal solutions of a convex formulation of optimization problem, Fig. 4(A,E)). The obtained solution is then approximated by ``string art'' generator \cite{Generator} according to \ref{a1}.   

In order to evaluate the quality of ``string art'' approximations of optimal solutions, we use two measures. The measure $\delta_{x}(\mathbf{x})$ gives the quality of approximation of material density distribution $\mathbf{x}$. It is defined as:

\begin{equation}
\delta_{x}(\mathbf{x}) = \frac{\left\Vert \mathbf{x}_{b}-\mathbf{x}\right\Vert _{2}}{\left\Vert \mathbf{x}_{b}-\mathbf{x}_{o}\right\Vert _{2}}
\end{equation}  

here $\mathbf{x}_b = \mathbf{I}f$ is the baseline material distribution -- uniform distribution with the given volume fraction; $\mathbf{x}_o$ is the optimal material distribution that solves \ref{e1}. By definition, $\delta_{x}(\mathbf{x}_{o}) = 1$, $\delta_{x}(\mathbf{x}_{b}) = 0$. 
In order to quantify the elastic properties of the given material distribution, we use another measure. It is given as

\begin{equation}
\delta_{c}(\mathbf{x}) = \frac{ c(\mathbf{x}_{b})-c(\mathbf{x})}{c(\mathbf{x}_{b})-c(\mathbf{x}_{o})}
\end{equation}  
 
where $c(\mathbf{x})$, $c(\mathbf{x}_{b})$ and $c(\mathbf{x}_{o})$ are strain energies computed according to \ref{e1} for the tested, baseline and optimal distributions, respectively. Again, by definition, $\delta_{c}(\mathbf{x}_{o}) = 1$, $\delta_{c}(\mathbf{x}_{b}) = 0$.

We approximate the obtained optimal solutions with the ``string art'' generator, using default number of pins - $N_p = 288$ and the line thickness $\alpha = 1$. Both the optimal, baseline solution and ``string art'' approximation are performed within the discrete domain of $600 \times 600$ pixels. For each test example, we produce three different ``string art'' approximations, using $4000$, $2000$ and $1000$ lines (Fig. 4(B-D),(F-H)). 

Both optimal solutions depicted in Fig. 4(A,E) are found for the volume fraction of $0.5$. ``String art'' generator \cite{Generator} does not provide control over integral brightness of the obtained pattern of lines, therefore, in order to provide a correct comparison of solutions, we perform balancing of a sting art solution, in order to make it correspond to a given volume fraction. It is done in the following way. If the total weight of all fibers exceeds $\mathbf{I}f$, the density of fibers is scaled to the value providing correct volume fraction. If the total weight of fibers is lower than $\mathbf{I}f$, the density of the matrix is increased to meet the correct volume fraction.   

\begin{table}
	\label{tab:benchmark1}
	\caption{MBB beam}
	\medskip
	\centering
	\begin{tabular}{|c|c|c|c|c|c|}
		\hline 
		Specimen & $N_{l}$ & $x$ & $c(x)$ & $\delta_{x}, \% $ & $\delta_{c}, \%$ \tabularnewline
		\hline 
		\hline
		Baseline &   & $x = 0.5$ & $57.4929$ & $0$ & $0$ \tabularnewline
		\hline 
		Optimized &   & $x \in [0,1]$ & $33.2643$ & $100$ & $100$ \tabularnewline
		\hline 
		Spec. 1 & 4000 & $x_{fibers} = 0.7016$, $x_{matrix} = 0$ & $46.2279$ & $54.00$ & $46.49$ \tabularnewline
		\hline 
		Spec. 2 & 2000 & $x_{fibers} = 0.9125$, $x_{matrix} = 0$ & $39.4459$ & $79.51$ & $74.49$ \tabularnewline
		\hline 
		Spec. 3 & 1000 & $x_{fibers} = 1.0$, $x_{matrix} = 0.1492$ & $36.2501$ & $82.75$ & $87.68$ \tabularnewline
		\hline 
	\end{tabular}
\end{table}

\begin{table}
	\label{tab:benchmark2}
	\caption{Cantilever beam}
	\medskip
	\centering
	\begin{tabular}{|c|c|c|c|c|c|}
		\hline 
		Specimen & $N_{l}$ & $x$ & $c(x)$ & $\delta_{x}, \% $ & $\delta_{c}, \%$ \tabularnewline
		\hline 
		\hline
		Baseline &   & $x = 0.5$ & $21.2312$ & $0$ & $0$ \tabularnewline
		\hline 
		Optimized &   & $x \in [0,1]$ & $14.7283$ & $100$ & $100$ \tabularnewline
		\hline 
		Spec. 1 & 4000 & $x_{fibers} = 0.6700$, $x_{matrix} = 0$ & $19.0693$ & $42.16$ & $33.25$ \tabularnewline
		\hline 
		Spec. 2 & 2000 & $x_{fibers} = 0.8831$, $x_{matrix} = 0$ & $18.0832$ & $66.37$ & $48.41$ \tabularnewline
		\hline 
		Spec. 3 & 1000 & $x_{fibers} = 1.0$, $x_{matrix} = 0.1200$ & $17.1362$ & $77.76$ & $62.97$ \tabularnewline
		\hline
	\end{tabular}
\end{table}

Tables 1 and 2 summarize the results of our experiments. In all cases ``string art'' structures approximating optimal solutions performed noticeably better than the baseline uniform material distributions. 

Composite structures with nonzero density of a matrix material worked particularly well, approaching $87.68\%$ and $62.97\%$ of the performance of optimal material distributions. Framed structures with zero matrix density performed somewhat worse, probably due to significant stress concentrations at joints - the feature that is known to limit the stiffness optimality of framed structures \cite{Sigmund2016}.

It is interesting to note that there is quite clear correspondence between $\delta_{x}(\mathbf{x})$ and $\delta_{c}(\mathbf{x})$ measures - good approximation of density distribution of an optimal solution is accompanied by the values of compliance close to optimal.

\section{Discussion}
\label{disc}

\subsection{Efficiency of ``string art'' approximation}

As we could see, the proposed phenomenological approach clearly works efficiently, at least for the case of cross-linked fibers studied above. The reasons for that are not straightforward. Indeed, the algorithm \ref{a1} does not guarantee the global optimality of the solution \ref{e2}. What is more important, even if we had a reliable way to approximate the material distribution \ref{e1} with poliline \ref{e2}, the approximation of density or stiffness does not guarantee the approximation of global elastic properties. One can demonstrate that vanishingly small change in local stiffness (say, a single column of pixels with small density) can have a dramatic effect on the global strain energy, while the global approximation of density will be nearly unaffected. 

In fact the distribution \ref{e2}, besides being a good approximation to optimal layout \ref{e1}, has another important property - it presents the framed structure, with frame rods mostly following the directions of rods found by topology optimization algorithm. This important property explicitly follows from the approximation procedure \ref{e2}. Such rods almost do not store bending and shear elastic energy, providing a good approximation of global property of minimum strain energy. In this respect our structures are somewhat similar to lamp shades made of thread \textit{papier mache}.

\subsection{Possible generalizations}

It is clearly possible to modify the optimization formulation \ref{e2} in such a way that it would explicitly account for sensitivities of elastic strain energy functional along the new string placement and not the difference in density distribution. However, even straightforward appropriation of conventional greedy ``sting art'' algorithm showcased here demonstrates impressive results and displays great potential of the suggested approach.

As has been mentioned above, the described modeling technique is good for representing composite materials with the cross-linked fibers. 
Nevertheless, for a wide class of loading situations with tensile/compressive loads acting on bundles of fibers dominating over shear/bending loads, the difference in elastic properties of a composite with and without cross-links is negligible. Moreover, a relatively straightforward modification allows addressing the problem of such materials with fibers without cross-links. In such a case one needs to use the anisotropic element stiffness matrices, with stiffness and anisotropy measure corresponding to homogenization of the material with differently oriented fibers. The development of such algorithm is of great practical importance and clearly of high priority for future work.   

The proposed approach can be straightforwardly generalized in few other ways. First, it can be  applied to the design of bent plates and curved shells, with only minor technological modifications. Figure 5 demonstrates conceptual possibility of production of a spherical surface using single coordinate ``string art'' machine. In this case fibers are not freely tensioned between pins, but laid onto a puncheon defining the shape of the resulting composite shell. The technology can be further generalized towards approximation of two-dimensional material distributions over shells of complex shapes and topologies, however, that would require significantly more complex manufactuing technology. Second, the simple technology of single-coordinate winding described above is suitable for applying prescribed tensile force to the fiber, producing composite materials with pre-strained reinforcement fibers. This is particularly important for civil engineering materials with low tensile strength of the concrete matrix, that can be significantly improved via pre-stressing. 
   
\begin{figure}
	\begin{center}
		\includegraphics[width=10cm]{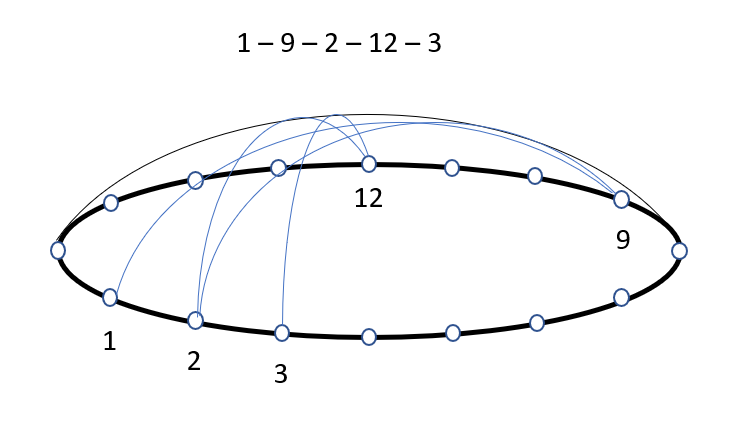}
		\protect\caption{Conceptual illustration a curved shell productuction using ``string art'' technology}
	\end{center}
\end{figure}

\subsection{Practical limitations of the approach}

It is clear that the major gain associated with our approach comes from replacement of the regular arrangement of fibers with seemingly disordered one, that does not allow close-packing of fibers. This constitute a serious problem and leads to a number of drawbacks. It is quite hard to control the plate thickness, random stress concentrations and inhomogeneities in a bulk of such a material. In order to ensure sufficient material's homogeneity, the thickness of the plate should be significantly larger than the thickness of a fiber. This introduces the constraints on the maximum volume fraction of fibers in a composite layer.      

\section{Conclusion}

In this work we first presented the idea of employing ``string art'' algorithms as the basis for the design of optimal composite materials. The idea has been demonstrated on the example of a rather idealized case of homogenization of two-phase composite material, produced by ``string art'' algorithms. This case roughly corresponds to the composite with cross-linked fibers. It was demonstrated that patterns generated by ``string art'' reproduce well the elasticity of optimized structures. 
The examples presented are rather illustrative -- as has been discussed above, there are multiple issues related to our methodology, that need to be addressed before it will reach a maturity sufficient for industrial applications. However, the major merit of our idea is that within our approach the optimal composite production does not need sophisticated technology to produce it - technological process of knitting the optimal structure can be successfully performed on a single degree of freedom winding machine. This ensures the possibility of mass production of such composites at scales sufficient, for example, for the commercial aircraft production.


\bibliographystyle{unsrtnat}
\bibliography{manuscript}

\end{document}